\documentclass[allclo]{FBSart}
\usepackage{amsmath,amsfonts,amssymb,graphicx}

\title{Spin dependent parton distributions and structure functions}
\author{W. Bentz\instnr{1}, I. C. Clo\"et\instnr{2}, 
T. Ito\instnr{1}, A. W. Thomas\instnr{3}, K. Yazaki\instnr{4}}
\instlist{Department of Physics, School of Science, Tokai University,
Hiratsuka-shi, Kanagawa 259-1292, Japan
\and Physics Division, Argonne National Laboratory, Argonne, 
IL 60439, USA
\and Jefferson Lab, 12000 Jefferson Avenue, Newport News, 
VA 23606, USA
\and Department of Physics, Tokyo Woman's Christian University,
Suginami-ku, Tokyo 167-8585, Japan}

\runningauthor{W.\,Bentz}
\runningtitle{Spin dependent parton distributions}
\begin{document}

\maketitle

\begin{abstract}
Nuclear parton distributions and structure functions are determined
in an effective chiral quark theory. We also discuss
an extension of our model to fragmentation functions. 
\end{abstract}

Effective chiral quark theories are powerful tools to
incorporate the nucleon quark substructure into the physics
of nuclei. For example, the EMC effect~\cite{Arneodo:1992wf} has shown that the
quark distributions of bound nucleons differ
from those of free nucleons, and this effect can be explained 
if one takes into account the response of
the quark wave function to the mean fields inside the nucleus~\cite{Cloet:2006bq}.
In this work, we concentrate on the model predictions for polarized
nuclear structure functions, and also
briefly discuss an extension to describe fragmentation 
functions~\cite{progress}.  

We will be concerned with the following EMC ratios:
\begin{eqnarray}
R(x) = \frac{F_{2A}(x_A)}{Z F_{2p}(x) + N F_{2n}(x)},
\qquad
R_s^H(x) = \frac{g^H_{1A}(x_A)}{P_p^H g_{1p}(x) + P_n^H g_{1n}(x)} \,.  
\label{ratios} 
\end{eqnarray}
Here $x$ is the usual Bjorken variable, and $x_A$ is $A$ times
the Bjorken variable for the nucleus of mass number $A$. The 
structure functions of the nucleon (nucleus) are denoted as $F_{2}$, 
$g_{1}$ ($F_{2 A}$, $g^H_{1 A}$ with $H$ the spin projection along 
the beam direction). 
$P_{\alpha}^H$ are the polarization factors of
protons and neutrons. Both ratios in Eq.~(\ref{ratios}) become unity 
in a naive nonrelativistic single particle model.

Usually only a few valence nucleons (or holes) contribute
to the nuclear polarization, and $g_{1A}^H$ is of order
$1/A$ relative to $F_{2A}$. Also,
the structure function of a proton is larger and better
known than that of the neutron. Therefore, possible
candidates for the observation of the polarized EMC effect are
stable nuclei which are not too heavy, and where the polarization 
is dominated by protons.

In our calculations, we describe the nucleon as a bound state of a 
quark and a diquark
by using the Faddeev framework in the Nambu--Jona-Lasinio (NJL) model
\cite{Ishii:1995bu}. We take into
account the scalar and axial-vector diquark channels, and include
confinement effects by eliminating
unphysical quark decay thresholds in the proper-time regularization
scheme \cite{Bentz:2001vc}. 

We calculate the unpolarized (polarized) quark distribution 
functions in the single nucleon by inserting the operator
$\gamma^+ \delta(x - k_-/p_-)$ ($\gamma^+ \gamma_5 \delta(x - k_-/p_-)$)
into the quark propagators which make up the nucleon Green function.
Finite nuclei in our present approach are described in a
simple independent particle picture by using depth parameters of the 
scalar and vector potentials from our earlier self-consistent
nuclear matter calculations \cite{Bentz:2001vc} and standard values for the range
and diffuseness. Using the Faddeev equation we translate these potentials
into the average fields for quarks, and use them in the calculation
of the quark distributions in a bound nucleon.
Finally, we calculate the momentum distributions
of the nucleons \cite{Cloet:2006bq}, and obtain the quark distributions 
in the nucleus and the nuclear structure functions by using the 
convolution formalism.

The resulting EMC ratios are shown in Figs. 1--3 for the nuclei
$^7$Li, $^{11}$B and $^{27}$Al.\footnote{The structure functions shown
in Figs.~1-3 refer to the leading multipoles ($K=0$ for the
unpolarized, $K=1$ for the polarized case), which are linear
combinations of the corresponding quantities in the helicity ($H$)
basis. For details, see Ref.~\cite{Cloet:2006bq}.} It is seen that the polarized EMC 
effect is predicted to be larger than the unpolarized one.
This corresponds to a reduction of the quark spin sum in the nucleus,
i.e., some amount of the quark spin is converted into orbital angular
momentum.

\begin{figure}[tbp]
\centering\includegraphics[scale=0.6]{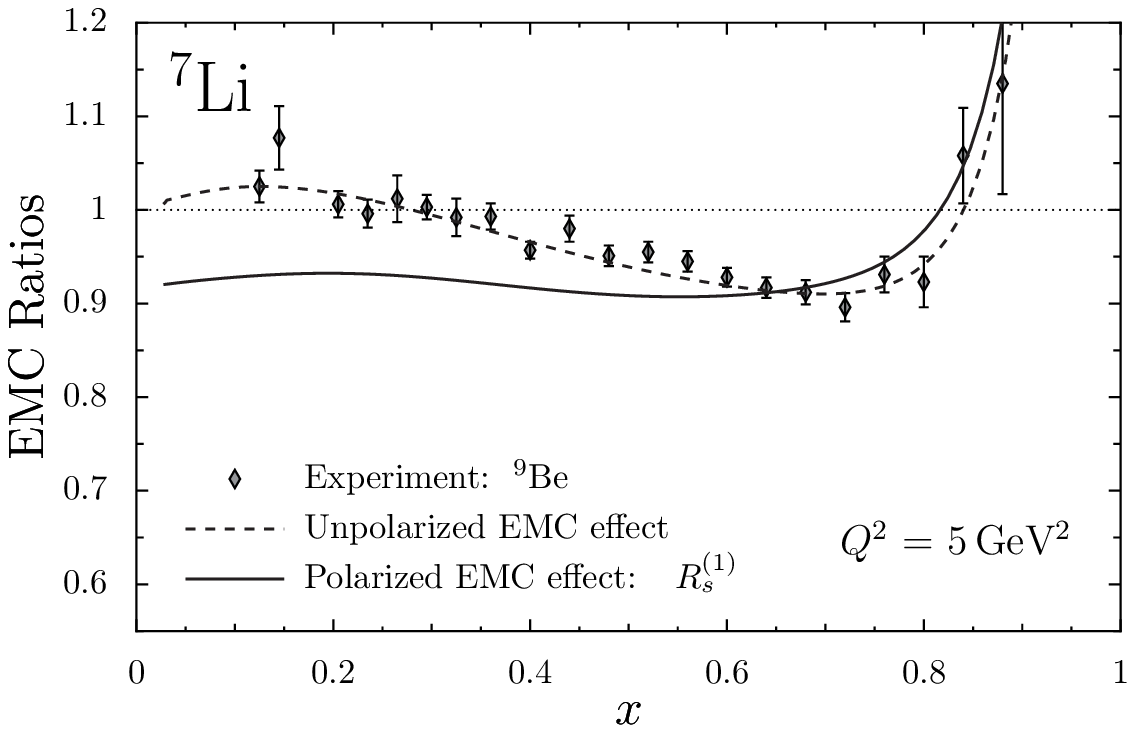}
\caption{EMC ratios for $^{7}$Li. The experimental data refer to $^{9}$Be.}
\end{figure}

\begin{figure}[tbp]
\centering\includegraphics[scale=0.6]{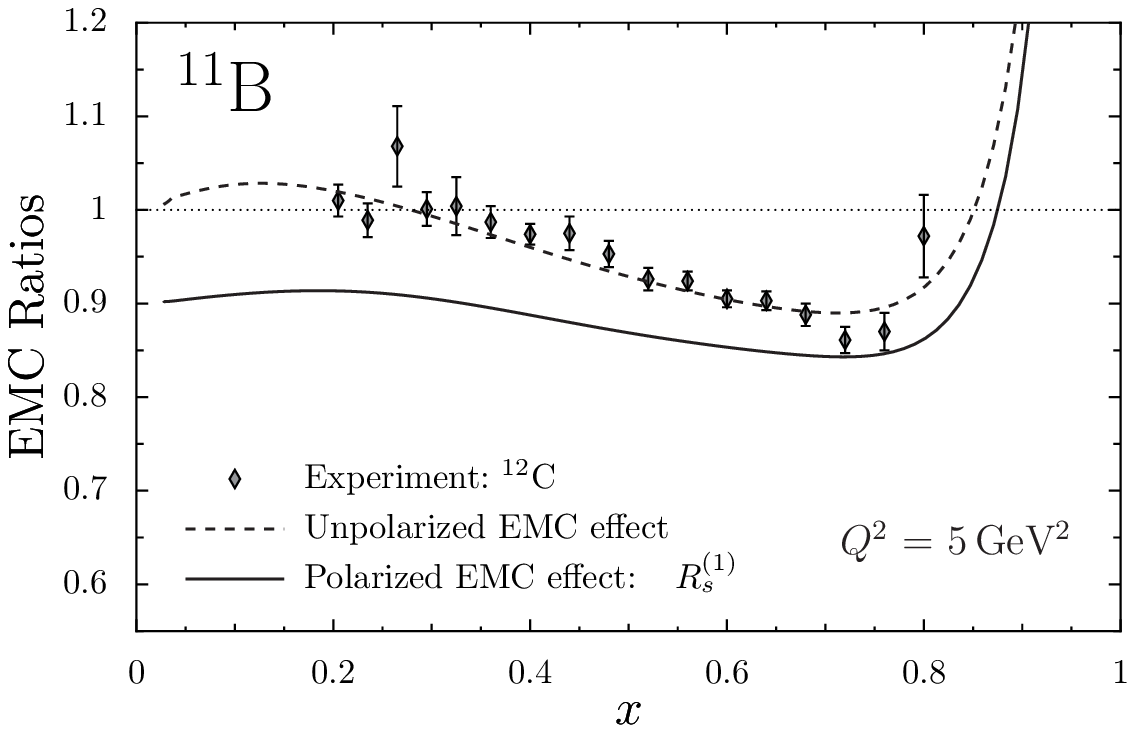}
\caption{EMC ratios for $^{11}$B. The experimental data refer to $^{12}$C.}
\end{figure}

\begin{figure}[tbp]
\centering\includegraphics[scale=0.6]{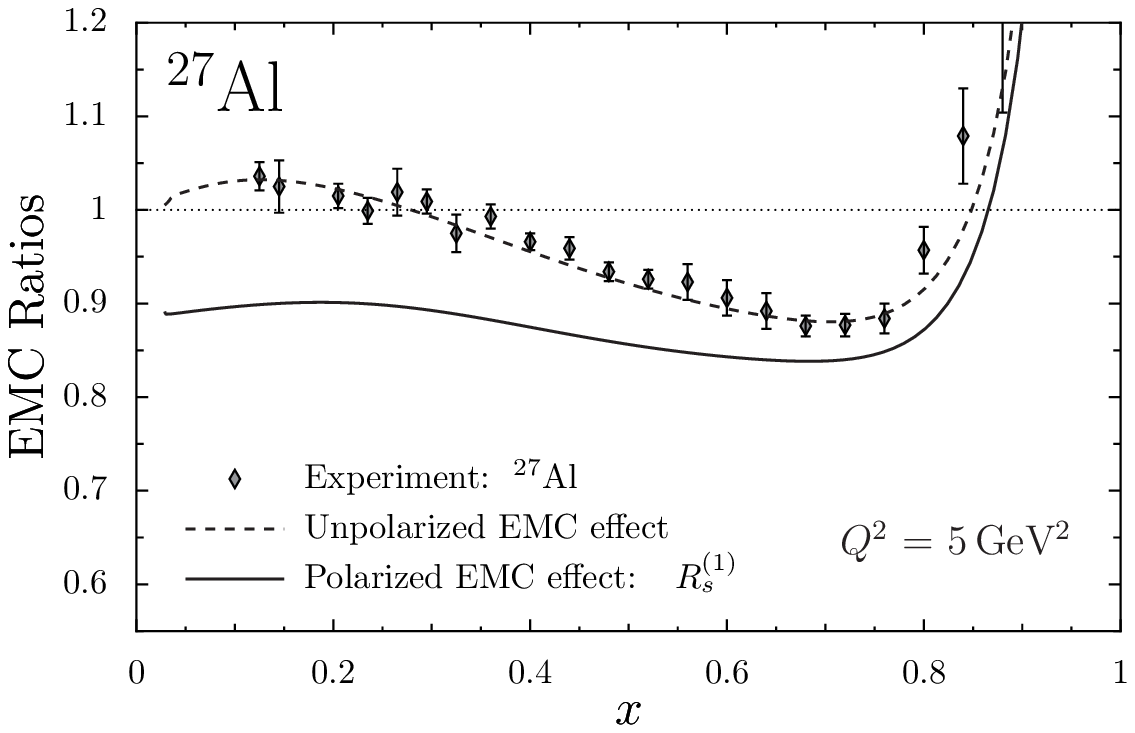}
\caption{EMC ratios for $^{27}$Al.}
\end{figure}

Here we comment on a possible extension of the model to fragmentation
functions: Starting from the operator definitions, one can use 
crossing and charge conjugation symmetries to show that
the distribution ($f(x)$) and fragmentation ($D(z)$) functions are 
essentially one and
the same function, defined in different regions of the variable. 
That is, if we define $f(x) = \Theta(1-x) F(x)$, then
\begin{equation}
D(z) = \pm \Theta(1-z)\, \frac{z}{6} \, F\left(x=\frac{1}{z}\right) \,, \nonumber
\end{equation}
where the plus (minus) sign holds if the hadron is a fermion (boson).
This relation is known in the literature as Drell-Levy-Yan
(DLY) relation \cite{Drell:1969jm}, and holds also in QCD up to leading
order. Numerical results and detailed
discussions based on this relation will be presented in a 
future publication \cite{progress}.

\begin{acknowledge}
The $Q^2$ evolutions were performed by using the NLO computer codes
of Ref.~\cite{Miyama:1995bd}. 
This work was supported by: Department of Energy, Office of Nuclear Physics,
contract no. DE-AC02-06CH11357, under which UChicago Argonne, LLC, operates
Argonne National Laboratory; contract no. DE-AC05-84ER40150, under which
JSA operates Jefferson Lab, and by the Grant in Aid for Scientific Research
of the Japanese Ministry of Education, Culture, Sports, Science
and Technology, project no. C-19540306.  
\end{acknowledge}


\begin{thebibliography}{9}








\bibitem{Arneodo:1992wf}
  M.~Arneodo,
  Phys.\ Rept.\  {\bf 240}, 301 (1994).

\bibitem{Cloet:2006bq}
  I.~C.~Clo\"et, W.~Bentz and A.~W.~Thomas,
  Phys.\ Lett.\  B {\bf 642}, 210 (2006).

\bibitem{progress} W. Bentz, T. Ito, A.W. Thomas, and K. Yazaki, to be published.

\bibitem{Ishii:1995bu}
  N.~Ishii, W.~Bentz and K.~Yazaki,
  Nucl.\ Phys.\  A {\bf 587}, 617 (1995).

\bibitem{Bentz:2001vc}
  W.~Bentz and A.~W.~Thomas,
  Nucl.\ Phys.\  A {\bf 696}, 138 (2001).

\bibitem{Drell:1969jm}
  S.~D.~Drell, D.~J.~Levy and T.~M.~Yan,
  Phys.\ Rev.\  {\bf 187}, 2159 (1969);
  S.~D.~Drell, D.~J.~Levy and T.~M.~Yan,
  Phys.\ Rev.\  D {\bf 1}, 1617 (1970).

\bibitem{Miyama:1995bd}
  M.~Miyama and S.~Kumano,
  Comput.\ Phys.\ Commun.\  {\bf 94}, 185 (1996);\\
  M.~Hirai, S.~Kumano and M.~Miyama,
  Comput.\ Phys.\ Commun.\  {\bf 108}, 38 (1998).

\end{thebibliography}
\end{document}